\documentstyle[psfig]{mn}
\def\H0{{\it H}$_0$}
\def\Ms{{\it M}$_\odot$}

\def\q0{{\it q}$_0$}

\def\ergps{erg~s$^{-1}$}
\def\kmpspMpc{km~s$^{-1}$~Mpc$^{-1}$}
\def\Ms{{\it M}$_\odot$}

\def\Zs{$Z_{\odot}$}

\def\ie{i.e., }

\title[Supernova heating] 
{The effect of supernova heating on cluster properties and
  constraints on galaxy formation models}
\author[K. K. S. Wu et al.] 
{\parbox[]{6.5in} {K. K. S. Wu,$^{1,3}$ A. C. Fabian$^1$ and
P. E. J. Nulsen$^2$}\\
\\
$^1$ Institute of Astronomy, Madingley Road, Cambridge CB3 0HA\\ 
$^2$ Department of Physics, University of Wollongong, Wollongong
     NSW 2522, Australia\\
$^3$ kwu@ast.cam.ac.uk}
\date{}

\begin{document}

\maketitle

\begin{abstract}
  Models of galaxy formation should be able to predict the properties
  of clusters of galaxies, in particular their gas fractions,
  metallicities, X-ray luminosity-temperature relation, temperature
  function and mass-deposition-rate function.  Fitting these
  properties places important constaints on galaxy formation on all
  scales.  By following gas processes in detail, our semi-analytic
  model (based on that of Nulsen \& Fabian 1997) is the only such
  model able to predict all of the above cluster properties.  We use
  realistic gas fractions and gas density profiles, and as required by
  observations we break the self-similarity of cluster structure by
  including supernova heating of intracluster gas, the amount of which
  is indicated by the observed metallicities. We also highlight the
  importance of the mass-deposition-rate function as an independent
  and very sensitive probe of cluster structure.
\end{abstract}


\section{Introduction}

Our understanding of galaxy formation through the hierarchical
clustering of gas and dark matter (DM) has improved greatly over the past
decade. The aim of models of galaxy formation is to account for the
properties of all structure, from the first dwarf galaxy to the rich
clusters of galaxies that are virialising now.  The collapse of the
dominant dark matter is relatively well understood, so that the major
current problem is to model the complex gas processes that take place
within collapsed and collapsing DM halos.  Despite their success in
accounting for some properties of galaxies, many semi-analytical
models do not predict correctly the properties of groups and clusters
of galaxies.  In particular, these models should predict gas
fractions, metallicities, the X-ray luminosity-temperature ($L_{\rm
  X}-T$) relation, the temperature function and the
mass-deposition-rate ($\dot{M}$) function of clusters.

One problem is that the gas fraction in clusters, which is observed to
be in the range 10--20 per cent (White et al. 1993b; White \& Fabian
1995; David, Jones \& Forman 1995; White, Jones \& Forman 1997; Evrard
1997\nocite{wnef93,wf95,djf95,wjf97,evrar97}), is underestimated by
those models that have low baryon fractions (e.g., most of the
values considered by Kauffmann, White \& Guiderdoni
(1993\nocite{kwg93}), Cole et al. (1994\nocite{cafnz94}), Heyl et al.
(1995\nocite{hcfn95}), Kauffmann \& Charlot (1998\nocite{kc98}) and Baugh et
al. (1998\nocite{bcfl98})).

In this paper we show that additional heat input to the intracluster
gas, which in our fiducial model we take to be due to supernovae, can
consistently account for the major X-ray properties of groups and
clusters of galaxies.  This has a similar effect to `preheating' as
invoked by other authors (Kaiser 1991; Evrard \& Henry 1991; Metzler
\& Evrard 1994; Navarro, Frenk \& White 1995; Bower 1997; Jones et al.
1998\nocite{kaise91,eh91,me94,nfw95,bower97,jsepw98}).  A well-known
problem addressed by preheating is the slope of the $L_{\rm X}-T$
relation. Models and simulations which do not involve preheating (and
in general those which investigate galaxy properties do not) are
generally `self-similar' and give $L_{\rm X}\propto T^2$ (Kaiser
1986\nocite{kaise86}), whereas clusters are observed to follow $L_{\rm
  X}\propto T^3$ (David et al. 1993). Models which preheat the gas in
clusters generally break the self-similarity. For example, a constant
excess specific energy can have a large effect in the shallow
potential wells of groups, while having little effect on the largest
clusters. The excess energy raises the temperature as well as the
potential energy of the gas by pushing it outwards, thereby flattening
the gas density profile. This lowers $L_{\rm X}$ and steepens the
relationship between $L_{\rm X}$ and $T$.

During galaxy formation gas is heated and possibly ejected from
galaxies by supernovae. It retains most of this `excess energy' as
thermal and gravitational potential energy (and possibly kinetic
energy) as it passes through the collapse hierarchy, eventually
becoming part of the intergalactic medium in groups and clusters.  Gas
that is ejected from a galaxy is expected to recollapse with the next
`major merger' (which in our model is the doubling, at least, of the
mass of the halo that the galaxy resides in). This is reasonable since
ejected gas would encounter the inter-galactic medium which
subsequently forms part of a larger halo. In our fiducial model the
only source of heating and metal enrichment is Type II supernovae (SNe
II) associated with star formation (Nulsen \& Fabian 1995,
1997\nocite{nf97,nf95}; hereafter NF95 and NF97). There is some
evidence that most of the metals in intracluster gas came from SNe II
rather than SNe Ia (Mushotzky et al.\ 1996\nocite{mlatf96}; Nagataki
\& Sato 1998\nocite{ns98}), but a further reason for our assumption is
because while both types inject about the same amount of energy, SNe
Ia inject about 10 times more iron than SNe II. Hence for the iron
abundance measured in intracluster gas, the associated energy
injection is maximised by using only SNe II.
We assume an energy injection of $4 \times
10^{50}$ erg per supernova (Spitzer 1978\nocite{spitz78}) which is
not radiated locally. The
mean mass of iron produced per supernova is 0.07 \Ms, which leads to
3.7 keV/particle excess specific energy in gas with solar iron
abundance. By choosing the initial metallicity in groups when they
are first formed to be a suitable value, the iron abundance of
clusters in our model is $Z\approx 0.3$\Zs\, which corresponds to
about 1.1 keV/particle of excess energy.

The semi-analytical model we use is based on that described in NF97,
with an improved cosmological model and a more accurate form for dark
matter halos.  The model pays particular attention to gas processes,
explicitly including the effects of cooling flows in its treatment of
hot gas. When hot gas has had sufficient time to radiate its thermal
energy it is assumed to preferentially form low mass stars,
effectively baryonic dark matter (BDM), as observed in cooling flows
in clusters. We assume that dark matter halos follow the NFW
profile (Navarro, Frenk \& White 1997\nocite{nfw97}; hereafter NFW97).
Gas in the resulting potential wells is assumed to be isothermal and
in hydrostatic equilibrium. The model clusters turn out to have gas
density profiles closely resembling those in observed clusters.

In this paper we use an open cosmology with $\Omega_0=0.3$, \H0 $=50$
\kmpspMpc and a cosmological constant of zero. Density fluctuations
are obtained from a CDM model with primordial spectral index of $n=1$,
normalised to make $\sigma_8=0.7$. This normalisation, which is
slightly lower than some estimates based on cluster numbers, is
discussed in section~3. We note that our results would be largely
unchanged in a flat cosmology with the same $\Omega_0$, since the mass
function of clusters today would be the same. Likewise, our
conclusions do not depend on when the excess energies were injected or
the exact history of the clusters and their progenitors. We do however
assume that the supernovae injected their energy into virialised gas,
as opposed to the uncollapsed IGM.

\section{Description of the model}
The two main changes from the model described in NF97 are the use of
an open cosmology and the improvement of the DM and gas profiles. The
mass of a halo is defined as that inside the virial radius, $r_{200}$,
and the mean density inside $r_{200}$ is required to be 200 times the
background density of an $\Omega=1$ universe of the same age. (This is
marginally different from NFW97, but follows the spherical
collapse model strictly.) We use the NFW profile (NFW97) which takes
the form
\begin{equation}
  \rho(r) = \frac{\rho_s}{x (1+x)^2},
\end{equation}
where $x = r/r_s$, and the characteristic density and scale, $\rho_s$
and $r_s$, are parameters. For a given mass virialising at some time
in a given cosmology (which yields $r_{200}$), $\rho(r)$ has only one
degree of freedom, which can be fixed by applying the algorithm given
in the Appendix of NFW97 to obtain the concentration, $c=r_{200}/r_s$.
$\rho_s$ is then fixed since we know the mean density inside $r_{200}$.
The algorithm used was shown by NFW97 to provide reasonable fits of
$c$ to simulation results, where small halos were found to be more
`concentrated' than large ones. In our cosmology the value of $c$
typically ranges from 4.5--6.5 for clusters.

We consider the above profile to describe the
total density of DM and gas, thus the gravitational potential
$\phi(r)$ is calculated from it, assuming that the halo is truncated
at $r_{200}$:
\begin{equation}
   \phi(r) = \alpha\left( -\frac{\ln(1+x)}{x} + \frac{1}{1+c} \right),
\end{equation}
where $ \alpha = 4\pi G \rho_s r_s^2$. Since the gas fraction is small,
the DM also follows the NFW profile to a good approximation.

In our model the gas is isothermal and in hydrostatic equilibrium,
giving the gas density profile
\begin{equation}
  \rho_{\rm g}(r) \propto \exp\left( -\frac{\mu m_{\rm H}}{kT} \phi(r) \right),
\end{equation}
where $\mu m_{\rm H}$ is the mean mass per particle in the gas, $T$ is its
temperature and $k$ is Boltzmann's constant.  Inserting the expression
for the potential gives
\begin{equation}
   \rho_{\rm g}(r) = A (1+x)^{\eta/x}
\end{equation}
where $A$ is a constant and $\eta=\mu m_{\rm H}\alpha/(kT)$.  The
above profile, which we shall refer to as the NFW-gas profile, was
shown by Makino et al. (1997\nocite{mss97}) to closely approximate the
conventional $\beta$-model (with $\beta=\eta/15$). It tends to $A
e^\eta$ in the centre and to $A$ for large $x$. Thus $A$ is, in
practice, very small and the plateau at large $x$ is not noticeable.
The parameter $\alpha$ has the same units as velocity dispersion
squared, thus it is apparent that $\eta$ is closely related to the
widely-used parameter $\beta$, the ratio of the DM
kinetic energy to the thermal energy of the gas when both are
isothermal.

The NFW-gas profile has been fitted to the surface brightness profiles
of observed clusters and it models most clusters very well (Ettori \&
Fabian 1998\nocite{ef98}).  Values of $\eta$ in the largest clusters which
have cooling flows (suggesting they are in a relaxed state, in contrast
to non-cooling flows; Buote \& Tsai 1996\nocite{bt96}) average
about 10.5, with a standard deviation of about 1.0 (Ettori, private
communication). Direct comparison shows that the gas profile for
$\eta=10.5$ closely follows the NFW profile from $x=0.4$ to 3. The
profiles differ inside $x=0.4$ due to the flat core of the NFW-gas
profile, which implies that the gas fraction increases as a function
of radius, as observed in real clusters (e.g.\ White, Jones \& Forman
1997\nocite{wjf97}).  This last effect becomes more pronounced and
spreads to larger radii for smaller values of $\eta$, the major cause
of which will be heating by supernovae.

Since $\eta$ gives both the gas temperature and the slope of the gas
density profile, it is the structural parameter through which
supernova heating exerts its influence on cluster properties. The procedure
for setting $\eta$ is as follows. We postulate that the total specific
energy of the gas (thermal plus gravitational) is proportional to the
specific gravitational energy of the DM, and that any excess specific
energy from supernova heating is added to the above gas energy, as its name implies.
\ie
\begin{equation}
  \frac{3kT}{2\mu m_{\rm H}}+
  \frac{\int\rho_{\rm g}\phi {\rm d}V}{M_{\rm gas}} =
   K \frac{\int\frac{1}{2}\rho_{\rm DM}\phi \mathrm{d}V}{M_{\rm DM}} 
               + E_{\rm excess},
\end{equation}
where $\rho_{\rm DM}$ is the DM density, which follows the NFW profile,
$M_{\rm gas}$ and $M_{\rm DM}$ are the total gas and DM masses, and
$E_{\rm excess}$ is the excess specific energy. The volumes of integration
are spheres of radius $r_{200}$.  Given the constant $K$, it is easy
to show that $\eta$ is only a function of $c$ and $E_{\rm excess}$. The
constant of proportionality $K$ is calibrated so that $\eta$ is
close to the observed value of 10.5 for the largest clusters,
when $E_{\rm excess}=0$.  We use the value $K=1.2$. (Note that
$\eta$ varies by around 0.5 depending on collapse redshift, through
its dependence on $c$.) The reason for matching to the largest clusters
is that if supernova heating does occur, we expect it to have least effect on
the largest clusters. As $c$ varies little over cluster masses, this
method ensures that clusters are self-similar when there is no
excess energy (\ie $c$ and $\eta$ are roughly constant).  The trends for
$c$ and $\eta$ are to increase slightly with decreasing mass, over all
masses in our model (see below).

The mass deposition rates of clusters are estimated by finding the
radius $r_{\rm cool}$ at which the cooling time is equal to the time
$\Delta t$ to the next major merger (see below) or the present, and
dividing the gas mass enclosed by $\Delta t$ to give $\dot{M}$. For
clusters with cooling flows, \ie those with significant $\dot{M}$, the
hot gas which cools does not form stars but BDM.

The Cole \& Kaiser (1988\nocite{ck88}) block model is used to simulate
the merging history of dark matter halos. The smallest masses in the
hierarchy are $1.5\times 10^{10}$\Ms\ and there are 20 levels of
hierarchy so that the total mass in one collapse tree is $7.9\times
10^{15}$\Ms. When a `block' in the hierarchy virialises, it merges
into one halo both older halos and previously uncollapsed material.
The new average metallicity and excess energy of the gas can therefore
be calculated. Since blocks double in mass up the hierarchy, each
collapse of a block roughly corresponds to a major merger. As already
mentioned, we set the metallicities in groups to be high without
considering the implications of this for star formation.  Halos with
masses in the range 15--120 $\times 10^{12}$\Ms\ are given initial
iron abundances of $Z=0.5$\Zs\.  (This implies that the gas in some of
the smaller halos is unbound.)  As a result the average metallicities
(for a given mass) range from 0.3\Zs\ for clusters of mass $250\times
10^{12}$\Ms\ to 0.25\Zs\ for the largest clusters. The range of scatter in
metallicities is $\approx 0.1$\Zs.
These are in good agreement with the recent {\it ASCA}
measurements by Fukazawa et al.  (1998\nocite{fmtex98}) of iron
abundances in 40 nearby clusters and the measurements of Allen \&
Fabian (1998\nocite{af98a}).

The gas fraction of the clusters in our model lie within one per cent
of 14 per cent.

\section{Results and Discussion}

\begin{figure}
\centerline{\psfig{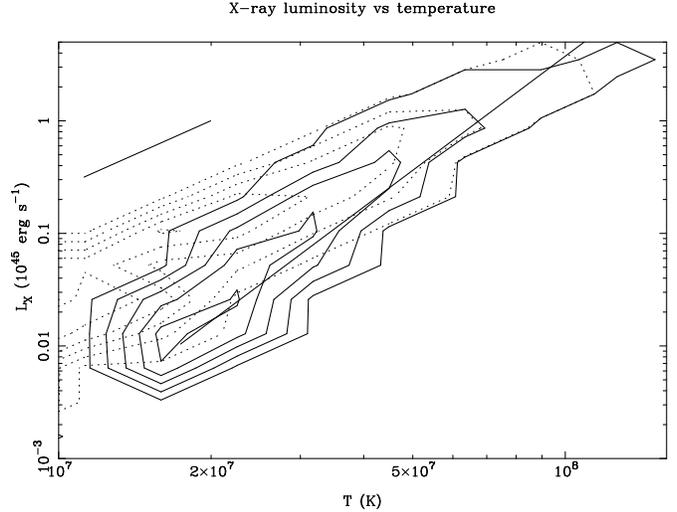}}
\caption{Contour plots of the luminosity-temperature distribution,
  without heating by supernovae (dotted lines) and with supernova heating
  (solid lines). The contours are spaced at equal logarithmic intervals. The
  long straight line is the best fit (for bolometric luminosities)
  taken from David et al.\ (1993). The distribution obtained without
  heating is roughly parallel to the short line segment which follows
  $L_{\rm X}\propto T^2$. In this case the smallest clusters are
  overluminous by an order of magnitude. The excess energies bring the
  smaller clusters into agreement with data.}
\label{lxt}
\end{figure}

To illustrate the effect of heating from supernovae, we display in
Fig.~\ref{lxt} the $L_{\rm X}$-$T$ distributions before
and after the inclusion of supernova heating. We superimpose on
the distributions the power law fit obtained by David et al.
(1993\nocite{dsjfv93}), for bolometric luminosities. The excess
energies bring the smaller clusters into agreement with the observed
correlation, by increasing their temperature, flattening their density
profiles and thereby decreasing their luminosities.  For clusters of
mass $2.5\times 10^{14}$\Ms, $\eta$ decreases to around 9.  The spread
in the $L_{\rm X}$-$T$ distributions are significant and arise naturally in
our simulation from the different formation histories of the clusters.

For comparison with Fig.~1, if we halve the excess energies injected,
the average luminosity of a $2\times 10^7$K cluster rises to $5\times
10^{43}$\ergps. Doubling the excess energies causes most of the
clusters with $T<3\times 10^{7}$K to be unbound, which clearly
disagrees with data, and roughly halves the luminosity of $3\times
10^{7}$K clusters. In future work we will investigate whether restricting
the heating to the core of clusters can decrease the excess energies
required (preliminary results suggest that this could reduce the total
energy requirements by about a third).

\begin{figure}
\centerline{\psfig{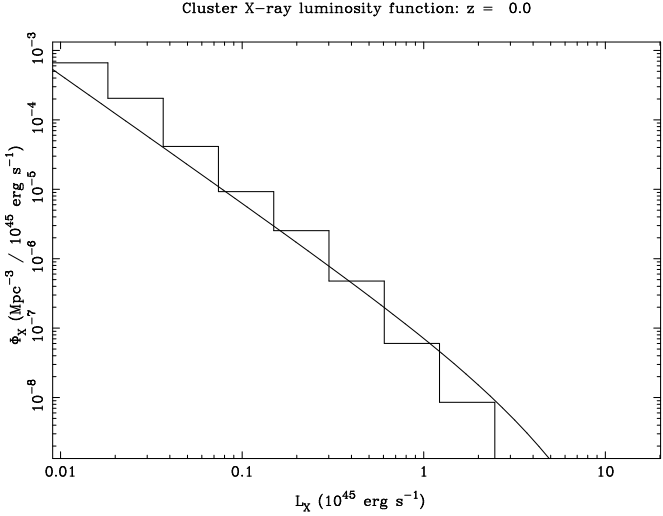}}
\caption{The X-ray luminosity function (XLF) from the model with
  supernova heating. The curve is the best fit Schechter function to the {\it
    ROSAT} Brightest Cluster Sample (BCS) bolometric luminosity
  function (Ebeling et al. 1997).}
\label{xlf}
\end{figure}

\begin{figure}
\centerline{\psfig{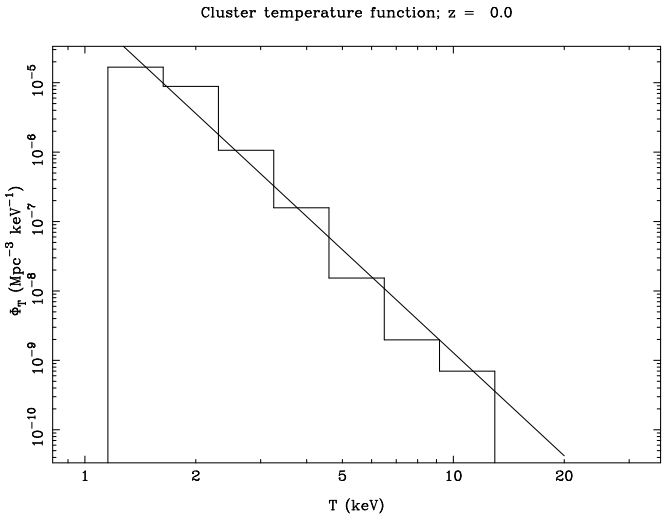}}
\caption{The temperature function from the model with supernova heating. The power
law fit to the temperature function obtained by Edge et
al. (1990) is plotted.}
\label{tf}
\end{figure}

In Figs.~\ref{xlf} and \ref{tf} we show the X-ray luminosity
function (XLF) and temperature function respectively. The curves
plotted for comparison are the best fit Schechter function to the {\it
ROSAT} Brightest Cluster Sample (BCS) bolometric luminosity function,
obtained by Ebeling et al. (1997\nocite{eefac97}), and the best power
law fit to the temperature function obtained by Edge et
al. (1990\nocite{esfa90}). The errors in the BCS XLF are very small,
while the temperature function is currently uncertain by factors of
around 2 or 3. A problem which follows from an
incorrect slope of the $L_{\rm X}$-$T$ relation is that if the temperature
function fits the data reasonably well, then the XLF is too steep compared
to data (e.g.\ Kitayama and Suto 1996\nocite{ks96}). Correcting the
$L_{\rm X}$-$T$ relation thus improves the slope of the XLF
significantly, though it is still slightly steeper than observed.

Estimates of $\sigma_8$ from cluster abundances in general use a
theoretically derived mass function and assume some mass-temperature
($M-T$) relation in order to compare with an observed temperature
function.  The calibration of the $M-T$ relation is thus crucial to
the results, especially as the temperature function has a very steep
slope.  We find that with zero excess energy our model clusters
obey the simple estimate (appropriate for isothermal $r^{-2}$ gas and DM
density profiles) of
\begin{equation}
  \frac{GM_{200}}{2r_{200}} = \frac{kT}{\mu m_{\rm H}} \label{mt}
\end{equation}
to within a few per cent, where $M_{200}$ is the total mass within
$r_{200}$. 
%(In the following discussion of the $M-T$ relation we
%consider for simplicity clusters which collapse at $z=0$ only, though
%the simulation naturally includes the formation history of clusters
%observed today.) 
Supernova heating raises the temperature
slightly for small clusters.  For a cluster
of mass $4.9\times 10^{14}$\Ms\ collapsing at $z=0$ 
the temperature increases from
$3.5\times 10^{7}$~K to $3.8\times 10^{7}$~K in the model described,
close to 10 per cent higher. The resulting temperatures in our model
are a necessary consequence of fitting the $L_{\rm X}-T$ relation.
In a recent paper, Eke et al. (1998\nocite{ecfh98}) used a more
elaborate form of Equation~\ref{mt} and included scatter in the M-T
relation, which brought down the predicted $\sigma_8$. With our CDM
shape parameter of $\Gamma=\Omega_0 h \exp(-\Omega_b
(1+1/\Omega_0))=0.12$ their Fig.~7 predicts a value of $\sigma_8\approx 0.75$.
Since preheating was not included, our value of $\sigma_8=0.7$ appears
consistent with their results.

\begin{figure}
\centerline{\psfig{figure=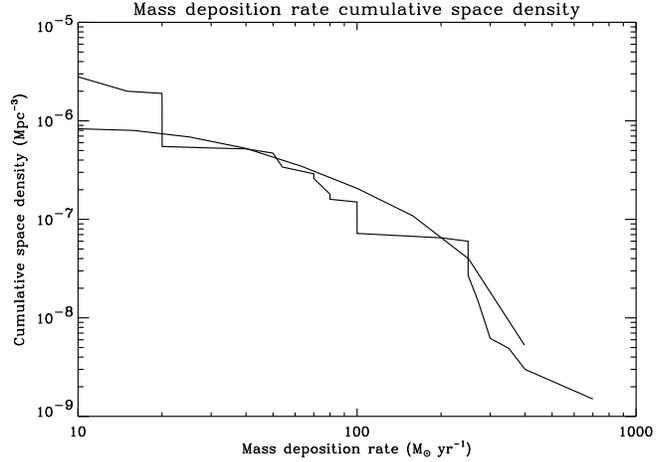,width=0.5\textwidth,angle=0}}
\caption{The smoother curve is the $\dot{M}$ cumulative function from
our model with supernova heating, and is compared to the function taken from
Peres et al. (1997) but with the age of clusters changed to 6 Gyr, as
explained in the text. We have also excluded clusters with $T<2\times
10^7$K; when all clusters are allowed to contribute, the function
increases below 50 \Ms\ yr$^{-1}$, rising to $6\times 10^{-5}$
Mpc$^{-3}$ at 10 \Ms\ yr$^{-1}$, but such clusters would not be
included in the observed sample.
}
\label{mdf}
\end{figure}

Turning to the mass deposition rates, we display the $\dot{M}$
cumulative space density in Fig.~\ref{mdf}. We have excluded
clusters with temperatures less than $2\times 10^7$ K since such
objects would be too faint in the 2--10 keV band to be included in the
sample used by the two papers below. Our result agrees very well with
the $\dot{M}$ cumulative function obtained by Edge, Stewart \& Fabian
(1992\nocite{esf92}) and lies slightly lower than that shown in Peres
et al. (1997\nocite{peres98}).  The systematic uncertainties in
measurements of $\dot{M}$ in clusters are at present around a factor
of 2. In particular the age of a cluster since its last major merger
is unknown and in determining $r_{\rm cool}$ the age of the universe is
generally used. Lowering this age in the analysis of Peres et al.
(1997) from 13 Gyr to 6 Gyr, which is the typical age of a 5 keV
cluster in our model, lowers $r_{\rm cool}$ and more than halves the
values of $\dot{M}$ in several clusters (Peres, private
communication). In Fig.~\ref{mdf} we show the cumulative function
from Peres et al. (1997) recalculated with this change in the age of
clusters. There is very good agreement for all but small values of
$\dot{M}$, which is not a serious concern since the clusters and their
measurements are both sensitive to small perturbations in this range.

The mass deposition rate of a cluster is a highly sensitive probe of
the gas density inside a radius of $\sim 0.1$ Mpc (the typical $r_{\rm
  cool}$ observed in cooling flows). On the
other hand, the luminosity in all but the most massive cooling flows
comes from inside a radius $\sim r_s$ which is larger than the cooling
flow region. Therefore $L_{\rm X}$ and $\dot{M}$ are independent
probes of the gas density profile at different radii, and due to their
sensitivity they are
efficient at eliminating models. For example, while we can adjust the
overall density inside $\sim r_s$ by changing the temperature, 
the ratio of the densities
inside the two radii described above have to be correct in order to
match both the $\dot{M}$ function and the XLF. We found that models
which used gas density profiles of the form $r^{-2\beta}$ (see NF97)
overpredicted the $\dot{M}$ function by an order of magnitude or more.
The agreement with the available data suggests that the NFW-gas
profile is much closer to the right shape.

%\begin{figure}
%\centerline{\psfig{figure=inject25-65.ps,width=0.5\textwidth,angle=270}}
%\caption{Contour plot of the mass deposition rate-luminosity distribution
%with supernova heating. The contours are spaced at
%logarithmic intervals. The inclusion of supernova heating brings the
%$\dot{M}$ of the smaller clusters down and into agreement with data.
%(Compare with scatter plot in Peres et al. 1997.)}
%\label{mdlx}
%\end{figure}

As a further example of the ability of $\dot{M}$ to discriminate
between models, without supernova heating we find an average value of
$\dot{M}\approx 100$\Ms yr$^{-1}$ for cooling flow clusters with
$L_{\rm X}=6\times 10^{43}$\ergps\, which is too high when compared
with Fig.~2 of Peres et al. (1997\nocite{peres98}).  However, heating
lowers the value to $\dot{M}\approx 30$\Ms yr$^{-1}$, which then agrees
with their data. On the other hand, large clusters with $L_{\rm
  X}=10^{45}$\ergps\ have the correct $\dot{M}$ with and without
heating. Using a hydrodynamic and N-body simulation, Katz \& White
(1993\nocite{kw93b}) modelled a Virgo-like cluster with cooling of gas
included, and obtained an excessively high value of $\dot{M}\sim
400$\Ms yr$^{-1}$. Our results suggest that non-gravitational heating
could simultaneously help solve such a problem.
%We show in
%Fig.~\ref{mdlx} the distribution of clusters in $\dot{M}$-$L_{\rm X}$
%space after the inclusion of supernova heating, to be compared with the
%corresponding scatter plot in Peres et al.\ (1997\nocite{peres98}).
The above results argue for flatter gas density profiles
in small clusters, independently of the $L_{\rm X}-T$ relation. We
consider it very promising that by correctly fitting the $L_{\rm X}-T$
relation, the predictions pertaining to $\dot{M}$ naturally agree with
the data.

\section{Conclusions}
We have presented a galaxy formation model in which the X-ray
properties of clusters agree well with what is observed.
The model is based on that described
by Nulsen \& Fabian (1997), and uses observed gas fractions and
metallicities. Realistic gas density profiles were used, which sit in
potentials given by the NFW profile. In particular, we obtained good
fits to the $L_{\rm X}-T$ relation, the X-ray luminosity function, the
temperature function and the $\dot{M}$ function, which we find to be
a very sensitive diagnostic of cluster structure in the core.

To resolve a number of problems, including getting the correct slope
for the $L_{\rm X}-T$ relation and avoiding excessive
mass deposition rates for low luminosity clusters,
self-similarity of clusters has to be broken.  In our model this is
achieved by flattening the gas density profiles of the smaller
clusters, which occurs naturally when we include heating of
intracluster gas due to the retention of energy from supernovae in all
previous collapse stages. We find that the energy required agrees
well with the supernova energy released in producing the observed
metallicities if most of this energy is retained in the intracluster
gas. The result is that the above
problems are simultaneously resolved once supernova heating is
included. 

We have highlighted the importance of the mass deposition rate and the
$\dot{M}$ function as diagnostics of cluster structure. The value of
$\dot{M}$ probes the gas density profile at smaller radii than $L_{\rm X}$
and hence is able to give us independent information. Due to the
sensitivity of quantities such as $L_{\rm X}$ and $\dot{M}$ to the structure
and density of the hot gas, and the steep slopes of the three
distribution functions we consider, detailed gas processes can be as
important as cosmological considerations in determining what we
observe.

\section*{Acknowledgements}
We thank Stefano Ettori for the results of surface brightness profile
fits, Clovis Peres for the new calculation of the $\dot{M}$ function,
and Vince Eke, Stefano, Clovis and Martin Rees for very helpful
discussions. KKSW is grateful to the Croucher Foundation for financial
support. ACF thanks the Royal Society for support. PEJN gratefully
acknowledges the hospitality of the Institute of Astronomy during part
of this work.

\bibliography{paper,bookastrophys}
\bibliographystyle{mnras}

\end{document}